\documentclass[conference]{IEEEtran}
\IEEEoverridecommandlockouts

\usepackage{cite}
\usepackage{amsmath,amssymb,amsfonts}
\usepackage{algorithm}
\usepackage{algorithmic}
\usepackage{graphicx}
\usepackage{textcomp}

\usepackage{xcolor}
\def\BibTeX{{\rm B\kern-.05em{\sc i\kern-.025em b}\kern-.08em
    T\kern-.1667em\lower.7ex\hbox{E}\kern-.125emX}}

\usepackage{tikz}
\newcommand{\circled}[1]{%
  \tikz[baseline=(char.base)]{
    \node[shape=circle,draw,inner sep=1pt] (char) {#1};
  }%
}

\usepackage{booktabs}  
\usepackage{multirow}  
\usepackage{xcolor}    
\usepackage{hyperref}  
\usepackage{graphicx} 
\usepackage{makecell}
\usepackage{array}

\usepackage{pifont} 
\usepackage{graphicx} 
\usepackage{booktabs} 
\usepackage{float} 
\usepackage{xcolor,colortbl}

\definecolor{green}{RGB}{200,255,200}  
\definecolor{red}{RGB}{255,200,200}    
\definecolor{yellow}{RGB}{255,235,156} 

\usepackage[most]{tcolorbox}

\begin{document}

\title{\vspace{0.32in}
Privacy-Preserving Federated Loss Exploration for\\
Vision-Based Perception under Non-IID Data%
\thanks{\copyright~2026 IEEE. Personal use of this material
is permitted. All other uses, including republication,
redistribution, resale, promotional use, or incorporation
into other works, require permission from IEEE.}
}

\author{
\IEEEauthorblockN{1\textsuperscript{st} Nergiz Yuca}
\IEEEauthorblockA{\textit{University of Passau, Germany}\\
nergiz.yuca@uni-passau.de}

\and

\IEEEauthorblockN{2\textsuperscript{nd} Christian Intern\`{o}}
\IEEEauthorblockA{\textit{Bielefeld University, Germany}\\
christian.interno@uni-bielefeld.de}

\and

\IEEEauthorblockN{3\textsuperscript{rd} Nikolay Matyunin}
\IEEEauthorblockA{\textit{Honda Research Institute Europe GmbH, Germany}\\
nikolay.matyunin@honda-ri.de}

\and

\IEEEauthorblockN{4\textsuperscript{th} Markus Olhofer}
\IEEEauthorblockA{\textit{Honda Research Institute Europe GmbH, Germany}\\
markus.olhofer@honda-ri.de}

\and

\IEEEauthorblockN{5\textsuperscript{th} Barbara Hammer}
\IEEEauthorblockA{\textit{Bielefeld University, Germany}\\
bhammer@techfak.uni-bielefeld.de}

\and

\IEEEauthorblockN{6\textsuperscript{th} Stefan Katzenbeisser}
\IEEEauthorblockA{\textit{University of Passau, Germany}\\
stefan.katzenbeisser@uni-passau.de}
}

\IEEEoverridecommandlockouts
\maketitle
\vspace{-0.18in}

\begin{abstract}

Vision-based image classification models benefit from Federated Learning (FL) across vehicles, however learning accurate models from distributed vehicular cameras is challenging due to data heterogeneity and privacy concerns. To improve performance on non-IID data, advanced FL optimization approaches have emerged that share auxiliary insights, such as client gradient behaviors. While these gradient-guided optimized FL methods
improve the model accuracy, their reliance on sharing additional gradient information introduces a previously unexplored privacy vulnerability. In our empirical privacy-leakage experiment, we show that a malicious server can infer clients’ label sets from shared guidance matrices under heterogeneous client label distributions (simulated via Dirichlet-based label partitions), between $50\%$ and $84\%$ accuracy. To address this privacy leakage, we propose \texttt{PPFLex} that replaces the central server
with three Secure Multi-Party Computation (MPC) servers to securely compute the global guidance matrix and perform federated averaging under semi-honest and malicious MPC server assumptions.
We evaluate \texttt{PPFLex} on widely used traffic-sign classification datasets (e.g., GTSRB and BTSC) as representative autonomous vehicle perception benchmarks, comparing against baselines. Our results show that \texttt{PPFLex} achieves accuracy comparable to the plain-text \texttt{FedLEx} pipeline while preventing privacy leakage, incurring acceptable MPC overhead.

\end{abstract}

\section{Introduction}

Connected and Automated Vehicles (CAVs) increasingly rely on machine learning to enable vision-based perception in intelligent transportation systems. In real-world deployments,  perception models are broadly employed to support training of image-based tasks within diverse automotive scenarios. These models are typically trained on large datasets,
allowing them to predict class labels for provided images~\cite{Kou2024pFedLVMAL}. Applications such as cooperative perception, driver monitoring, object detection and tracking, and traffic sign recognition contribute to Advanced Driver Assistance Systems (ADAS), and depend on large-scale visual data collected from distributed vehicular sensors operating in heterogeneous environments \cite{Chellapandi2023FederatedLF, yuca2025survey}.

In our work, we focus on one representative image-based perception task, Traffic Sign Classification (TSC), to demonstrate the applicability of our framework, while emphasizing that the proposed approach can be compatible with other
vision-based
automotive perception tasks. TSC enhances driving safety by categorizing road signs and is broadly integrated into ADAS and autonomous driving systems~\cite{Nidamanuri2022APR,kesbhat2025TrafficSD}. It requires reliable classification under varying lighting conditions, weather changes, and viewing angles. As traffic signs serve as key map features for navigation, accurate sign classification also supports localization and motion planning \cite{Ertler2019TheMT}.
However, the scale of vehicular data raises
communication overhead and
data protection challenges. Traffic sign distributions might reveal the vehicle’s driving profile, 
exposing recurring routes or location-based behavior patterns. Several works~\cite{Hossein2025ArgosAD, Liang2024FederatedLW,Padaria2023TrafficSC} propose Federated Learning (FL) to address these concerns by enabling collaborative training without sharing raw data and supporting 
data protection regulations such as the GDPR~\cite{Truong2020PrivacyPI}.

Despite its promise, FL is affected by challenges associated with data heterogeneity, as clients' data are often Non-Independent and Identically Distributed (Non-IID) due to imbalances and perturbations \cite{zhu2021federated}. In TSC, non-IID effects arise since vehicles operate across diverse regions and experience different sign distributions and environmental conditions. For example, some cars travel on highways, trucks often drive through mountains or rural areas, and buses usually operate on city roads~\cite{Lian2024TrafficSR}. Traditional server-side aggregation, FedAvg~\cite{McMahan2016CommunicationEfficientLO}, can be ineffective under 
heterogeneity, since multiple local updates can diverge from the global objective and reduce generalization across clients \cite{Collins2022FedAvgWF}. To tackle the limitations of standard averaging in these non-IID settings, researchers have 
adopted knowledge transfer techniques~\cite{Shao2023ASO} to harmonize learning across diverse client distributions. 

Federated Loss Exploration (\texttt{FedLEx})~\cite{FedLEx} is one such method that integrates guided transfer learning into FL
by analyzing gradient behavior to guide the learning process. Unlike conventional FL approaches, \texttt{FedLEx} does not rely on weight-sharing alone. Instead, it focuses on aggregating the gradient information from the different loss surfaces of each client. This approach is designed to address the challenges of non-IID data, where class distribution differences between clients can create biases that hinder learning performance.

A further concern is privacy: although FL provides a layer of protection, it remains vulnerable to inference attacks, where adversaries reconstruct private training data from intermediate model updates~\cite{Chai2023ASF,Lyu2020PrivacyAR,Mothukuri2021ASO}. Such attacks, including gradient and model inversion,
can expose sensitive client information, undermining FL’s privacy guarantees~\cite{DeepLF}. To address these concerns, several privacy-preserving technologies have been proposed. Homomorphic Encryption (HE)~\cite{Doan2023ASO} allows computations on encrypted data; however, it introduces high computational overhead.  Differential Privacy (DP)~\cite{Ji2014DifferentialPA} adds noise to protect privacy, though this may degrade model performance. Secure Multi-Party Computation (MPC)~\cite{Lindell2020SecureMC} allows a group of untrusted parties to process their data jointly without revealing individual information and is emerging as a promising solution for multi-party settings.
 However, existing MPC-based secure aggregation protocols are mainly focused on traditional FL averaging and do not focus on providing privacy for sharing additional
 knowledge that can be used to optimize performance in non-IID settings. On the other hand, while  \texttt{FedLEx} enhances traditional FL training by leveraging gradient analysis, it lacks privacy guarantees, as gradient deviations can still reveal sensitive information such as local label distributions, similar to traditional FL vulnerabilities. 

This paper addresses the resulting gap between (i) non-IID optimization methods that rely on additional shared gradient information (e.g., \texttt{FedLEx}) and (ii) secure aggregation techniques primarily designed for standard averaging. We propose \texttt{PPFLex}, a privacy-preserving FL framework for traffic sign classification that leverages MPC to enable the private computation of \texttt{FedLEx}'s guidance information. In traffic-sign classification, local label distributions reflect the road environments where vehicle data is collected. For instance, warning signs such as animal crossings or school crossings denote specific surroundings. Moreover, traffic-sign systems are not fully standardised across countries, and datasets are often country-specific~\cite{GmezSerna2018ClassificationOT}. For this scenario, we conduct privacy leakage and \texttt{FedLEx} accuracy evaluations on two datasets, GTSRB~\cite{gtsrb} and BTSC~\cite{belgium}, under non-IID client partitions.
Note that, in this paper, we study traffic sign classification as a representative use case within an automotive perception pipeline. The traffic sign detection problem (small-object detection at distance under motion glare, blur, and adverse weather) is out of scope for this work.

\textbf{Our contributions.} The main contributions of our work are:
\begin{itemize}
  \item We demonstrate that the additional gradient information shared in \texttt{FedLEx} introduces a privacy vulnerability that can leak clients' privacy-sensitive data and quantify its severity in a realistic setting. Our evaluation is performed within the traffic sign classification domain by using the GTSRB and BTSC datasets.
  
  \item We develop \texttt{PPFLex}, an end-to-end privacy-preserving implementation of \texttt{FedLEx} via MPC. Unlike existing methods that separately address either non-IID optimization or generic secure aggregation, \texttt{PPFLex} combines both, preserving model accuracy across multiple vehicles, under non-IID data in a privacy-preserving way.

  \item We study the trade-offs among model accuracy, privacy, and communication/computational overhead across different MPC protocols and adapt to different security assumptions, including semi-honest and malicious servers.
\end{itemize}

\section{Related Work}
\label{related}

This section reviews prior work on secure aggregation in federated learning, optimization methods for handling non-IID client data, and FL in traffic sign classification. 

\subsection{Secure Aggregation in FL}
 Various MPC-based secure aggregation protocols enable FL clients to share their locally trained models with a group of servers in a privacy-preserving manner. For example, SAFELearn~\cite{Fereidooni2021SAFELearnSA}
 framework employs
 fully homomorphic encryption and MPC
 for global model privacy. SafeFL~\cite{Gehlhar2023SafeFLMF}
 evaluates the performance of various FL techniques against privacy inference and poisoning attacks. ELSA~\cite{Rathee2023ELSASA} is a secure aggregation protocol for FL that assumes malicious security
 in a two-server setting.  WW-FL~ \cite{wwfl} introduces a unified MPC-based framework for large-scale FL, enabling global model privacy and robustness against poisoning attacks. SPEFL~\cite{spefl}
 implements FL for the Internet of Things (IoT), providing device privacy and robustness
 on resource-limited devices. Finally, ScionFL~ \cite{scionfl}
 proposes secure quantized aggregation method that reduces
 server-client communication in FL.
 Although these works ensure secure aggregation in FL,
 they do not address
 optimization challenges
 caused by non-IID data distributions, which our work targets.
 
\subsection{Optimizing FL on Non-IID Data}
Recent FL optimization studies address non-IID challenges through transfer learning and knowledge distillation. In this context, transferring information from a global server or other peers
to local clients can enhance client performance on unseen data\cite{tan2022federated, nguyen2022where}.
%
%
Furthermore, researchers apply
hierarchical clustering to 
categorize clients based on weight update dynamics for more effective grouping and tailored optimization \cite{9207469, 10114536730383673151}. Parallel efforts identify data distributions that more accurately reflect client diversity \cite{neth2025feddiversetacklingdataheterogeneity}.
Finally, relevant to \texttt{FedLEx}~\cite{FedLEx}, advances in adaptive gradient methods
such as AdamW \cite{loshchilov2019decoupledweightdecayregularization} and second-order optimizers are gradually replacing traditional stochastic gradient descent in FL. Techniques like group normalization
further accelerate convergence \cite{wu2018group,ruiz2025group}. 
Note that no existing work has yet leveraged loss function geometry transfer to guide client gradients integrated with MPC, a gap our research aims to fill.

\subsection{FL in Traffic Sign Classification}

Several works study distributed and federated training approaches for traffic sign classification and detection. Stergiou et al.~\cite{Stergiou2022AFL} design a LeNet-5–based distributed training framework and analyze how client count and communication rounds affect accuracy and edge-device resource usage in an autonomous driving setting, showing that decentralized CNN training
can achieve accurate recognition. Argos et al.~\cite{Hossein2025ArgosAD} propose a decentralized FL framework for traffic sign detection in CAVs, where vehicles train lightweight local detectors on partitioned traffic-sign classes and aggregate model parameters using
FedAvg and FedProx~\cite{Li2019FederatedLC}. Lian et al.~\cite{Lian2024TrafficSR} study communication-efficient training, where each client sends only its most important model update components (Top-k selection) to reduce communication costs and uses Adam-based~\cite{Kingma2014AdamAM} local training to improve convergence. Liang et al.~\cite{Liang2024FederatedLW} propose a heterogeneity-aware approach based on data-free knowledge distillation, where clients share only class-level aggregated intermediate representations and predictions. Padaria et al.~\cite{Padaria2023TrafficSC} present a combined split and federated learning framework for traffic sign classification on GTSRB under a 5G-enabled setting. 
In contrast to prior TSC-FL work that primarily targets accuracy, efficiency, or architecture design, \texttt{PPFLex} specifically targets privacy risks introduced by non-IID optimization values
and addresses them through MPC.

\section{Background: Federated Loss Exploration}
\label{FedLEx}

In a traditional federated learning setting,  clients train models on their local device, decoupling training from direct access to raw training data. Each client $i$ ($1 \leq i \leq K$) holds data  distributed according to $p_i(x, y)$ with $
D_i = \{(x_i^{(1)}, y_i^{(1)}), \dots, (x_i^{(N_i)}, y_i^{(N_i)})\}$, where  $x \in \mathbb{R}^d$ represents a $d$-dimensional feature vector and $y \in \{1, \dots, M\}$
corresponds to
a class label from $M$ categories. The global objective function represents the optimization goal,
and is defined to minimize the aggregate loss across all clients:

$$
\mathcal{L}(W_{\text{global}}) = \frac{1}{K} \sum_{i=1}^{K} \mathbb{E}_{(x_i, y_i) \sim p_i} [\mathcal{L}(W_{\text{global}}, x_i, y_i)],
$$
where $\mathcal{L}(W_{\text{global}}, x_i, y_i)$ denotes the global loss
function
for a data point $(x_i, y_i)$ with $\mathbb{E}_{(x_i, y_i) \sim p_i}$ denoting the expected loss over client's local distribution $p_i(x, y)$. A well-known FL aggregation scheme, Federated-Averaging (FedAvg)~\cite{McMahan2016CommunicationEfficientLO} optimizes this objective through iterative training rounds. First, the server distributes the initial global model
$W_{\text{global}}$ to clients. Then, clients perform local training on their data and send their updated model parameters back to the server. The server averages the updates to compute the global model. 

\texttt{FedLEx}~\cite{FedLEx}  extends this process
to handle the challenge of non-IID data in FL. In such settings, client data distributions differ, causing
slow global model convergence. To overcome this, \texttt{FedLEx} 
 introduces a ``loss exploration phase" that identifies which model parameters are most sensitive to the statistical differences in each client's data, thereby providing a consistent global model. 
We now briefly describe the main phases in \texttt{FedLEx}; for full details, we refer to the original paper~\cite{FedLEx}.

\textbf{Federated Loss Exploration.} In this phase, each client explores its loss landscape, identifying critical parameter directions via gradient deviations \cite{nikolic_guided_2023}. At its core, clients train the initial model on their local data and measure the magnitude of change for each model parameter, estimating its sensitivity to the client's data distribution. 
Using Adam~\cite{Kingma2014AdamAM} optimizer during exploration, the update rule
for the $m$-th model parameter $W_{i,m}$ for client $i$ at exploration step $e$ is:$$W_{i,m}^{(e+1)} = W_{i,m}^{(e)} - \eta\, \nabla \mathcal{L}_i\left(W_{i,m}^{(e)}; D_i\right),$$where $\eta$ is the learning rate and $\nabla \mathcal{L}_i(\cdot)$ is the gradient of the local loss function $\mathcal{L}_i$ for client $i$
for 
its local dataset $D_i$. The
update over a set number of exploration steps, $E_{\text{exp}}$, is:

\begin{equation*}
    W_{i,m}^{\text{Final}} - W_{i,m}^{\text{Initial}} = -\eta \sum_{t=1}^{E_{\text{exp}}} \nabla \mathcal{L}_i\left(W_{i,m}^{(t)}; D_i\right),
\end{equation*}
where $W_{i,m}^{\text{Initial}}$ and $W_{i,m}^{\text{Final}}$ are the parameter values at the beginning and end of the exploration phase,
and $t$ indexes the training steps.
Subsequently, the gradient variability of the loss landscape for each parameter $m$ is computed as the squared deviation: $ \delta_{i,m}^2 = \left(W_{i,m}^{\text{Initial}} - W_{i,m}^{\text{Final}}\right)^2$.
Finally, each client $i$ compiles these squared deviations into a local guidance matrix, $G_{\text{local}_i}$, where the $m$-th element is defined as $G_{\text{local}_{i,m}} = \delta_{i,m}^2$.

\textbf{Global Guidance Matrix Construction.} After the exploration phase, the server aggregates the local guidance matrices ($G_{\text{local}}$) from all clients and normalizes them to construct the global guidance matrix, $G_{\text{global}}$. Each entry in this matrix indicates the average gradient deviation for a specific model parameter across the clients.

 To construct $G_{\text{global}_m}$, we first compute $\bar{G}_m = \frac{1}{K} \sum_{i=1}^{K} G_{\text{local}_{i, m}}$, the average of local matrices across all parameters.
 Each entry $G_{\text{global}_m}$ indicates the average
 deviation of the gradient for the $m$-th parameter across clients. 
 Then we apply min-max normalization to compute the global guidance per parameter $m$ across all clients:

\begin{equation}
\label{global}
G_{\text{global}_m} = \frac{\bar{G}_m - \min\bigl(\bar{G}_m\bigr)}
                      {\max\bigl(\bar{G}_m\bigr) - \min\bigl(\bar{G}_m\bigr)}.
\end{equation}

Here $\min\bigl(\bar{G}_m\bigr)$ and $\max\bigl(\bar{G}_m\bigr)$ denote the minimum and maximum values across all entries of $\bar{G}$. Finally, the guidance matrix  \(G_{\text{global}}\) is sent back to the clients in order to guide their model updates in each round of FL.

\textbf{Federated Learning.}
Subsequently, the local training process begins with selected clients computing their gradient updates $\Delta W_{i,m}$ for each parameter $m$ of client $i$'s model. Instead of sending this update directly, \texttt{FedLEx} incorporates the global guidance matrix to modulate it:$$\Delta W_{\text{modulated}_{i,m}} = \Delta W_{i,m} \times G_{\text{global}_ {m}},$$where $\Delta W_{\text{modulated}_{i,m}}$ is the
update for the $m$-th parameter.
 Finally, clients send these
 updates to the server, which then applies federated averaging to update the global model.

Overall, \texttt{FedLEx}~\cite{FedLEx} improves model convergence in non-IID settings through the additional gradient exploration and exchange of local matrices. In the next section, we describe how this FL procedure can be implemented more securely, to prevent a malicious server from inferring sensitive data from gradient and model updates.

\begin{figure*}[t!]
\centering
\includegraphics[trim=3.2cm 5.3cm 3.4cm 6.1cm,clip,width=\textwidth]{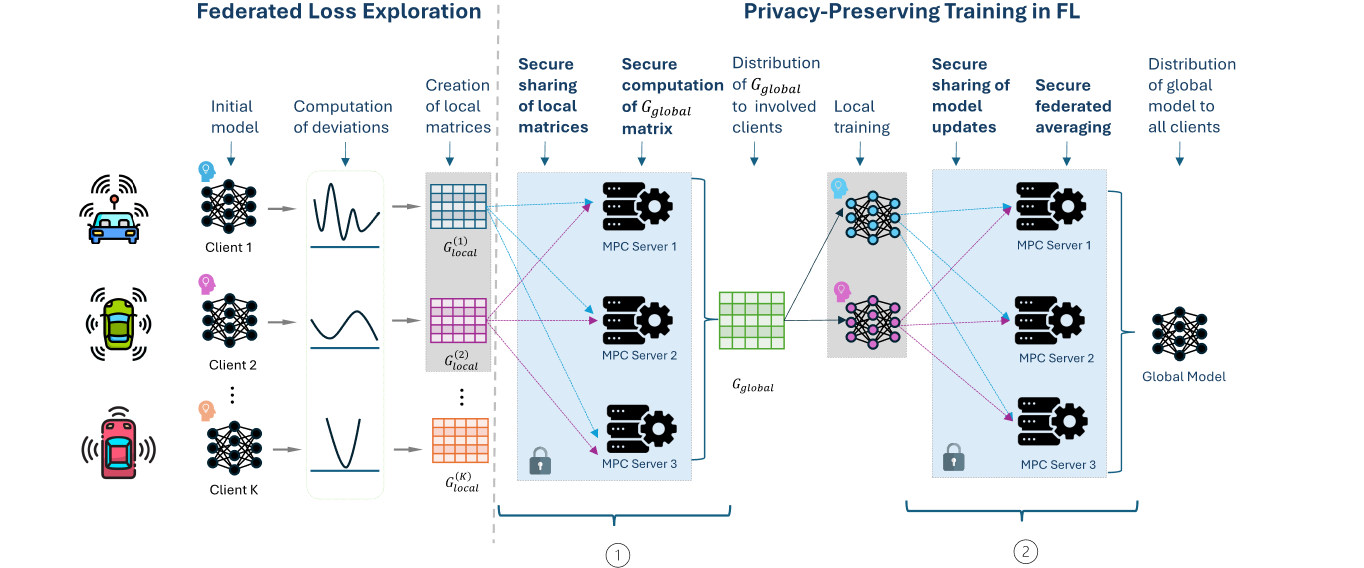}
\caption{PPFLex: MPC-based privacy-preserving federated learning in traffic sign classification.
In the exploration phase (Step~1), each client computes a local guidance matrix 
$G^{(i)}_{\text{local}}$ from parameter deviations and secret shares it among three MPC servers, 
which jointly compute the aggregated $G_{\text{global}}$.
In the training phase (Step~2), $G_{\text{global}}$ is distributed to clients to guide local updates, 
which are again secret-shared for secure
federated averaging.}
\label{fig:pncarchitecture2}
\end{figure*}

\section{Proposed PPFLex Framework}
\label{proposed}

 Our \texttt{PPFLex} framework uses MPC to provide privacy guarantees in \texttt{FedLEx}. As an application domain, we target FL for traffic sign classification, where vehicles' label distributions are naturally non-IID, and model deviations can reveal sensitive information about local driving environments. To address privacy vulnerabilities associated with single-server FL aggregation, we replace the centralized server in \texttt{FedLEx} with three MPC servers to securely compute $G_{\text{global}}$ (Eq.~(\ref{global})) and perform federated averaging in a privacy-preserving manner. Generally, the proposed  architecture
 supports two or more servers. The three-server default setting is selected to evaluate various underlying MPC protocols, including honest-majority protocols designed for three parties (Replicated2k, PSReplicated2k), while also supporting dishonest-majority protocols (SPDZ2k, Semi2k). Our approach ensures that the model updates and additional gradient information are shared with  MPC servers without revealing
client data.  The proposed enhancements to \texttt{FedLEx} are illustrated in Figure \ref{fig:pncarchitecture2}.

In our work, we employ \emph{secret sharing–based} MPC protocols. Their  core underlying primitive is secret sharing, which enables splitting a secret~$s$
among~$n$ parties such that any subset of size~$t+1$ can reconstruct
the secret, while any
$t$ or fewer parties learn no information.
Linear
operations, such as addition and scalar multiplication, can be performed locally on the
shares, whereas multiplications of shares rely on preprocessing and interaction among the servers.
In our setting, each vehicle decomposes its input into three shares, sending one share to each server. \textcolor{black}{
The servers jointly compute
$\mathbf{G}_{\text{global}}$ and perform federated averaging on secret-shared values, without reconstructing any raw client inputs.}  

As an application scenario, we consider a fleet-learning deployment in which vehicles periodically contribute to
a traffic sign classification using local camera data. Vehicles act as FL clients, while FL training is coordinated
and managed by the model owner (e.g., an Original
Equipment Manufacturer (OEM) or mobility provider). In this use case, class distributions correlate with geographic and behavioral factors, as traffic sign types and frequencies differ across countries, regions, and road infrastructures. A malicious server observing exploration updates may therefore infer location information, which \texttt{PPFLex} is specifically designed to prevent via MPC.

\texttt{PPFLex} involves two main MPC computations corresponding to the two privacy-sensitive server-side phases in \texttt{FedLEx}:

\textit{1) Global guidance matrix.} At the beginning of the 
learning round,
 randomly selected participating clients send 
their gradient deviations (computed locally in the exploration phase) to the three MPC servers, in a secret-shared format.
As the first MPC computation in \texttt{PPFLex} framework, the MPC servers
aggregate them to compute $G_{\text{global}}$ for involved clients using Eq.~(\ref{global}) (see \circled{1} in Figure~\ref{fig:pncarchitecture2}). 

\textit{2) Federated Learning.} After a local training phase, selected participating clients send their model updates to the three MPC servers in a secret-shared format. As the second MPC computation in \texttt{PPFLex} framework, the servers perform secure aggregation of updated model parameters of contributing clients to construct a global model (see \circled{2} in Figure~\ref{fig:pncarchitecture2}).

\textbf{Threat Model.}
The primary privacy objective is to prevent the server from learning client-specific properties (e.g., which traffic-sign classes appear in a vehicle’s local data), apart from what can be inferred from the global model outputs. For both MPC computations, we consider two security settings. In the semi-honest setting, servers follow the protocol but attempt to infer additional information about an individual’s input. In the malicious setting, servers may arbitrarily deviate from the protocol. 
Note that while \texttt{PPFLex} prevents server-side leakage through secure MPC computation, it does not defend against malicious clients
(assumed to be honest in this work). Integrating aggregation algorithms robust to malicious clients can be addressed in future work.
We evaluate \texttt{PPFLex} under both semi-honest and malicious settings in Section~\ref{Eval}.

Let $n$ denote the number of MPC servers and $t$ the maximum number of servers that may be corrupted and collude. For honest-majority replicated protocols, we assume $t < \frac{n}{2}$. In our default deployment with $n = 3$, this corresponds to at most one corrupted server. For dishonest-majority \texttt{SPDZ2k}, security holds as long as at least one computation party remains honest.
We assume three non-colluding MPC servers operated by independent entities (e.g., OEM backend, a Tier-1 mobility provider, and an independent service  provider). Clients (vehicles) are assumed semi-honest and do not collude with servers. The MPC servers receive only secret-shared data, without access to raw  
per-client gradient deviations or model updates. The global guidance matrix and global model are revealed as required to continue the learning process. Our implementation includes four protocols that operate over the arithmetic ring domain of integers modulo a power of two (Mod $2^{k}$), where $k$ denotes the bit length of the integers. In our work, we evaluate MPC protocols offering semi-honest (Semi2k, Replicated2k) or malicious (SPDZ2k, PSReplicated2k) security, depending on the assumed majority setting~\cite{mp-spdz}.

\textbf{Communication Optimization.}
Exploration-derived local guidance values $G_{\text{local}_i}$ are reused across multiple training rounds. Therefore, clients can upload them once, and the servers can store the corresponding secret shares. Subsequent secure guidance computations can then reuse the stored shares, avoiding repeated client uploads. This optimization
does not require reconstructing any client's local guidance in plain-text.

\section{Implementation} 
\label{implementation}

We implement the MPC-based outsourced secure FL aggregation pipeline, \texttt{PPFLex}, using the MP-SPDZ~\cite{mp-spdz} framework, 
which executes MPC programs written in Python-like code
and supports over $40$ protocols.
It allows us to evaluate the solution under different security assumptions and settings.


To address the privacy leakages that we identified in \texttt{FedLEx}, we implement
MPC programs 
for both the federated averaging and the guidance matrix computation.  Our implementation incorporates two optimizations. First, to minimize MPC communication overhead, the final normalization step for the guidance matrix is performed locally by each client,
avoiding costly non-linear min/max operations under MPC.
Second, we use a unified secure aggregation program that handles both the gradient updates and $G_{local}$ matrices and stores the results in the filesystem.
The source code is publicly available at https://github.com/nergiz-ue/PPFlex.

To integrate FL pipeline with implemented MPC programs, we make targeted adjustments to the \texttt{FedLEx} source. First, we dump MPC inputs (model updates and local matrices) to files and convert them  into fixed-point integers. Then, we replace plain-text executions with external calls to run MPC clients and computing parties.
This
approach allows us to seamlessly replace plain-text functionality with MPC in the existing FL  pipelines and evaluate MPC impact on the FL performance.

\begin{table}[t!]
    \centering
    \caption{Experiment Hyperparameters.}

    \begin{tabular}{l l}
        \toprule
        \textbf{Parameter} & \textbf{PPFLex} \\
        \midrule
        Dataset, Model & GTSRB / BTSC, 2CNN  \\
        Layers, hidden layers & 2, 200 \\
        Num. of epochs in Exploration  & 150 \\
         Batch size,  Learning rate  & 32, 0.0003 \\
        Optimizer  & Adam \\
       Loss function & CrossEntropyLoss \\
       Num. clients, total   & 20 \\
       Num. clients, participating  & 2 \\
       Num. numerical params (2CNN): &  1.68M\\
           \midrule
         Num. of MPC servers &  3  \\
           MPC protocols &   Semi2k, Replicated2k \\ & SPDZ2k, PsReplicated2k \\
           Ring size (all protocols) & $64$ bits (mod $2^{64}$) \\
Fixed-point precision & $24$ fractional bits\\
Mixed-domain conversions & daBits \\
    
        \bottomrule
    \end{tabular}
   \label{tab:FedLEx_params}
\end{table}

\section{Evaluation}
\label{Eval}

\begin{figure}[t]
    \centering
    \includegraphics[width=0.85\linewidth]{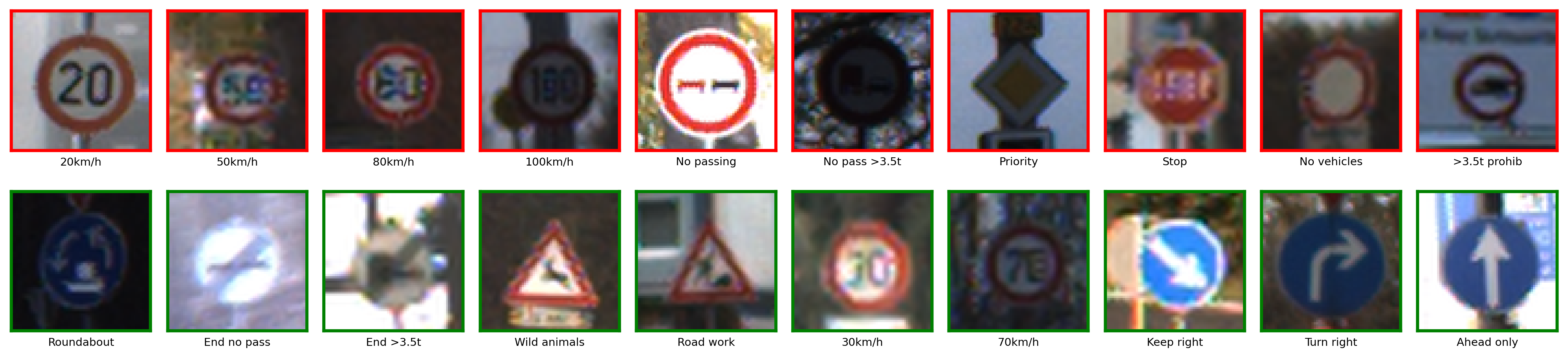}
    \caption{The example traffic sign images from the GTSRB dataset appear in different environmental settings.}
    \label{fig:class_examples}
\end{figure}

In this section, we evaluate the \texttt{PPFLex} framework along four dimensions: (i) model performance under non-IID TSC data, 
(ii) privacy leakage from \texttt{FedLEx} guidance information, 
(iii) the impact of MPC
on model accuracy, and 
(iv) MPC overhead.
%
We use the German Traffic Sign Recognition Benchmark (GTSRB), containing $26{,}640$ images across 43 traffic sign classes
(with examples shown in Figure~\ref{fig:class_examples}).
%
In addition, we use the Belgium Traffic Sign Classification dataset (BTSC), which contains $4{,}591$ training images and $2{,}534$ test images across 62 subclasses.
We include BTSC to evaluate cross-dataset generalization under distribution shifts. 


We employ the TwoCNN architecture with two convolutional layers followed by two fully connected layers, totaling 1{,}681{,}899 trainable parameters. Input images are resized to $32 \times 32 \times 3$. Unless stated otherwise, we use the Adam optimizer for both exploration and local training.
For FL experiments, local
training runs for 5 epochs with batch size 32 with learning rate 0.0003. We use different numbers of clients across experiments: $K = 20$ with $C=0.1$ for FL accuracy to assess convergence behavior, $K=10$ with 2 participating clients per round for MPC overhead,
and $120$ simulated clients for the privacy leakage experiment to provide sufficient statistical reliability for the inference evaluation.
To model heterogeneous client data, class distributions are generated using a Dirichlet distribution with parameter $\alpha$.
For MPC performance measurements, we run three MPC servers on \textit{e2-highmem-2} virtual machines (2 vCPUs, 16GB RAM each) on Google Cloud infrastructure in a 10\,GB LAN setting. The hyperparameter configuration follows the original
\texttt{FedLEx} setup~\cite{FedLEx} and is
summarized in Table~\ref{tab:FedLEx_params}.

\subsection{FedLEx Performance under Non-IID Traffic Sign Data}

First, we
compare \texttt{FedLEx} and FedAvg
under heterogeneous data distributions. Data heterogeneity is controlled using a Dirichlet distribution with $\alpha \in \{0.1, 0.5, 1.0\}$, where a smaller $\alpha$ indicates more skewed local class distributions. In all
experiments, we use $K{=}20$ clients, a participation rate of $C{=}0.1$, $E{=}5$ local epochs, batch size $B{=}32$, and the Adam optimizer with learning rate $0.0003$ and decay factor $0.99$. Each configuration is repeated over 5 independent runs, and we report the mean accuracy along with $\pm 1$ standard deviation.

\begin{figure}[t]
\centering
\includegraphics[trim=0.25cm 0cm 0cm 0cm,clip,width=1.02\columnwidth]{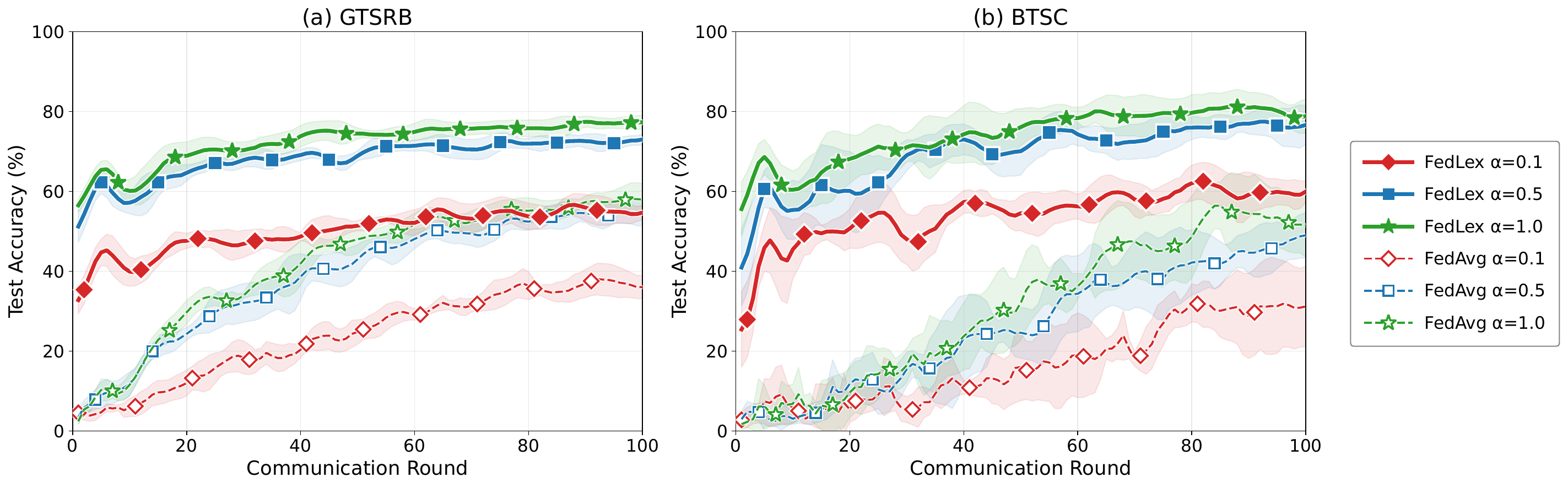}
      \caption{Test accuracy convergence of \texttt{FedLEx} vs FedAvg across Dirichlet $\alpha$ levels on (a) GTSRB and (b) BTSC datasets. \texttt{FedLEx} outperforms FedAvg in all settings.}      
\label{fig:alpha}
\end{figure}

As shown in Fig.~\ref{fig:alpha}, \texttt{FedLEx} consistently outperforms FedAvg across both datasets and all heterogeneity levels.
Notably, the performance gap widens as data heterogeneity increases. At $\alpha{=}0.1$, \texttt{FedLEx} outperforms FedAvg by an absolute margin of $18.66\%$ on GTSRB and $28.75\%$ on BTSC. Furthermore, \texttt{FedLEx} shows lower variance across runs, indicating more stable training under non-IID conditions.

\subsection{An Analysis of Privacy Leakage in FedLEx}
\label{leakage}

To empirically demonstrate the privacy vulnerabilities in \texttt{FedLEx}'s
pipeline,
we conduct
an experiment to evaluate whether a malicious server can infer clients' dominant traffic sign class from their transmitted guidance matrices. We simulate 120 clients, each holding data from a random subset of 1 to 4 classes. Within each client's class set, the relative proportions are drawn from a Dirichlet distribution.
Each client's local dataset contains 300 samples from the
respective dataset. Each client starts from a shared baseline 2CNN model and creates a local copy. The deviation is trained for 5 epochs with a learning rate of $0.01$, and a batch size of $32$.
After training, we compute the guidance vector as element-wise squared difference between the
updated and the original parameters.

\begin{figure}[t!]
    \centering
    \includegraphics[width=0.74\columnwidth]{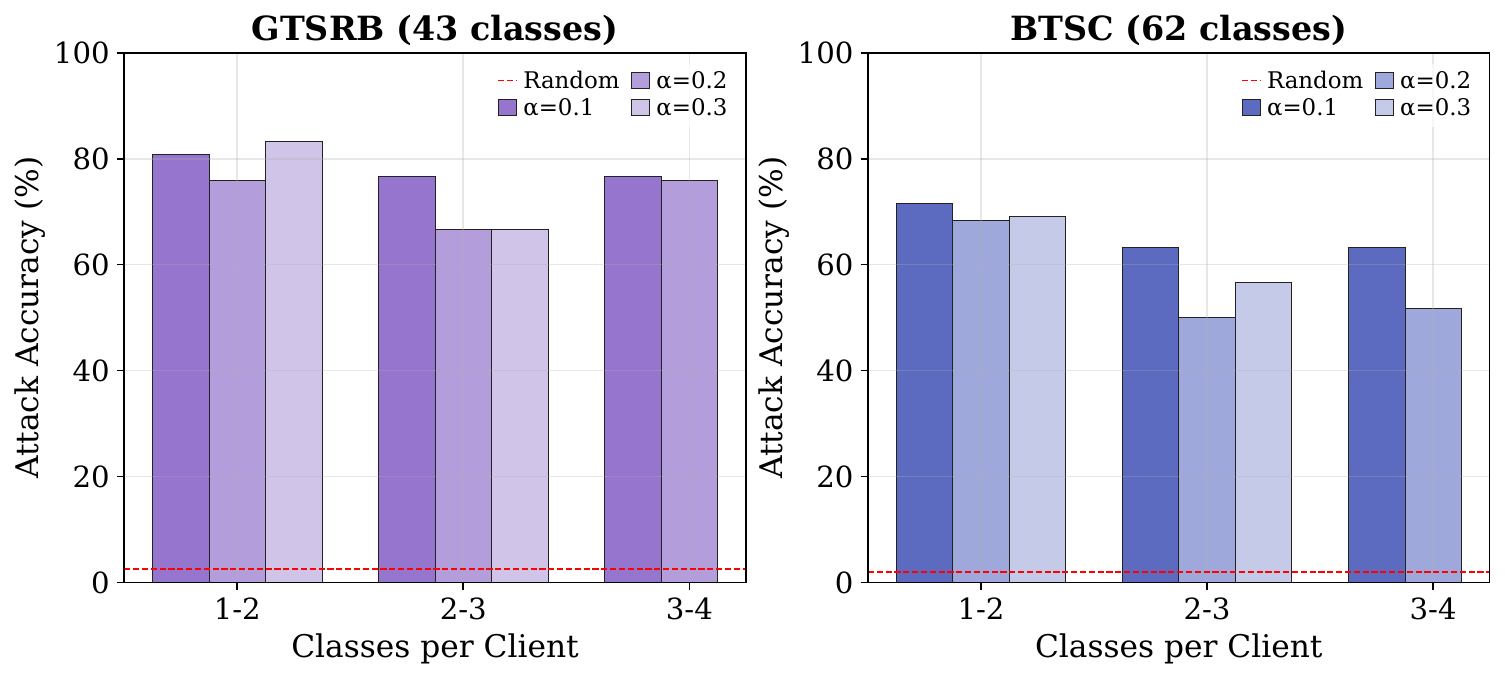}
    \caption{Attack accuracy for predicting a client's dominant traffic sign class from gradient deviations, grouped by the number of classes per client and data heterogeneity on GTSRB (left, 43 classes) and BTSC (right, 62 classes). 
    }
\label{fig:privacy_attack_heterogeneity}
\end{figure}

We then train a Random Forest classifier (100 trees, maximum depth 20) to predict each client's dominant class from this feature vector, using an 80/20 train/test split for evaluation. Prediction accuracy is measured using 5-fold cross-validation. We compare against a random guessing baseline, which yields approximately
$2\%$ accuracy given the large number of possible 
classes in the GTSRB and BTSC datasets.
As shown in Fig.~\ref{fig:privacy_attack_heterogeneity}, the attack achieves between $50\%$ and $84\%$ accuracy depending on the configuration.
Accuracy is highest when clients hold fewer classes, since the dominant class is more clearly reflected in the model update. Among the heterogeneity levels, moderate concentration ($\alpha{=}0.2$) yields the highest attack accuracy, while $\alpha{=}0.1$ and $\alpha{=}0.3$ produce comparable but lower results.
The results
also indicate 
that the privacy risk is not
specific to a particular dataset.
We note that the presented
attack cannot be launched under \texttt{PPFLex}, 
since no   single
server has access to raw client's inputs.


\subsection{MPC Impact on Model Accuracy }
\begin{figure}[t]
\centering
\includegraphics[trim=0.0cm 0.2cm 0.1cm 0.2cm, clip,width=0.85\columnwidth]{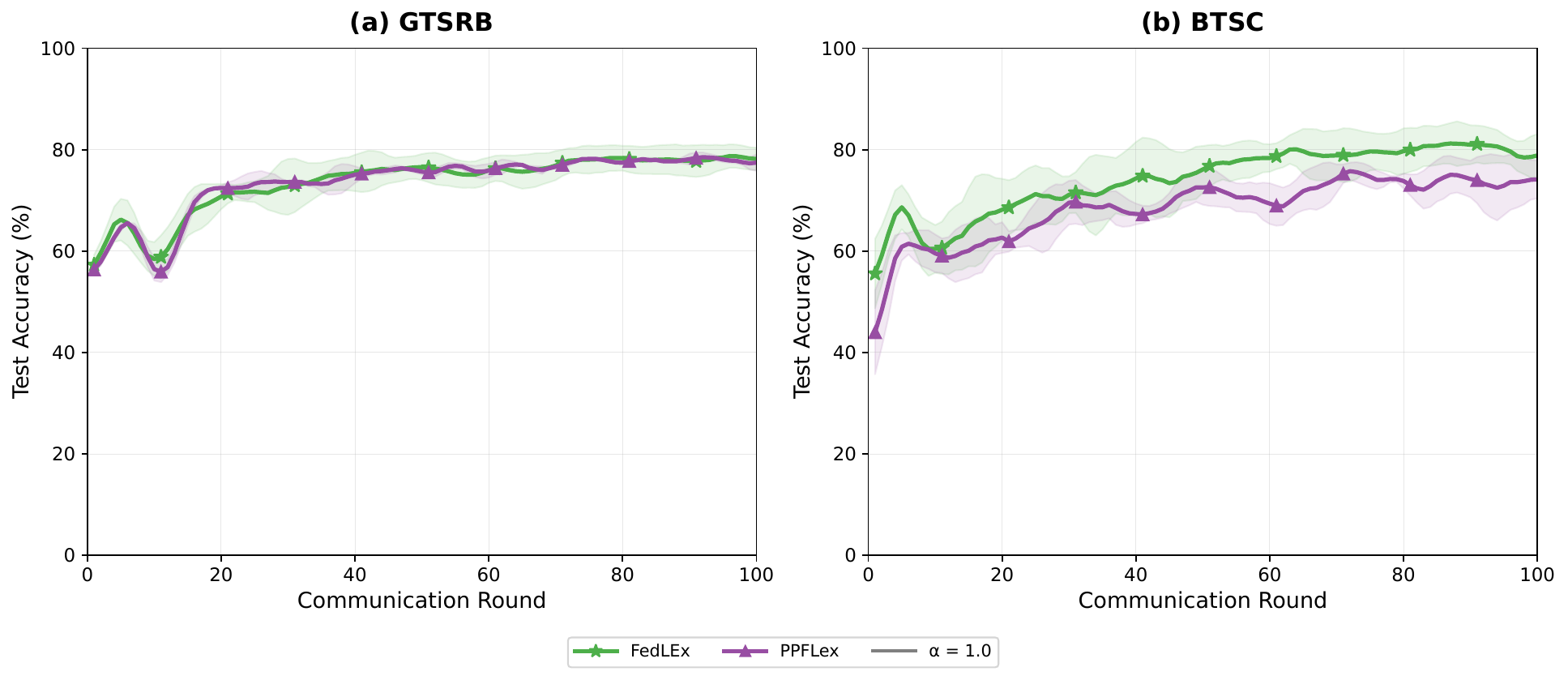}
\caption{Performance of PPFLex, in comparison to the plain-text \texttt{FedLEx}.}
\label{fig:ppflex-vs-fedlex}
\end{figure}

To verify that MPC extensions do not affect the FL performance (as MPC is done on
quantized integer values), we run both the original (unsecured) and MPC-enhanced FL pipeline
and compare the global model convergence 
after 100 training rounds. The results are averaged
across 5 runs
in both settings and presented in Figure~\ref{fig:ppflex-vs-fedlex}. 
The convergence curves of \texttt{FedLEx} and \texttt{PPFLex} closely align, with small discrepancies observed  
between the two,
attributed to the quantization of floating-point operations under MPC.
Overall, we confirm that MPC additions do not affect the overall FL performance and implement the desired functionality correctly.

\subsection{MPC Overhead}

\renewcommand{\arraystretch}{1.12}
\begin{table*}[ht]
\centering
\caption{
Average MPC overhead for secure federated computations (FL Averaging and Guidance Matrix) using the 2CNN model on GTSRB.
All protocols use 3~parties over $\mathbb{Z}_{2^k}$.
Shares generation denotes the approach of generating secret shares and their type.
Offline includes preprocessing and input supply; online includes computation and output reveal.
}
\label{tab:protocol-results}
\newcolumntype{G}{>{\columncolor{gray!20}}r}
\begin{tabular}{l c c l l | r r | G G | G }
\toprule
\textbf{MPC Protocol} & \textbf{\# Parties} & \textbf{Domain} & \textbf{Security} & \textbf{Shares Generation}
& \multicolumn{2}{c|}{\textbf{Offline}} & \multicolumn{2}{c|}{\cellcolor{gray!20}\textbf{Online}} & \textbf{Total} \\
& & & & & \textbf{Time (s)} & \textbf{Comm (MB)} & \cellcolor{gray!20}\textbf{Time (s)} & \cellcolor{gray!20}\textbf{Comm (MB)} & \textbf{Time (s)} \\
\midrule
Replicated2k  & 3 & $\mathbb{Z}_{2^k}$ & HM/SH  & Client (replicated)  & $\approx$0 & $\approx$0 & 20.4 & 26.9  & 20.5 \\
Semi2k        & 3 & $\mathbb{Z}_{2^k}$ & DM/SH  & Client (additive)    & $\approx$0 & $\approx$0 & 25.4 & 53.8  & 25.5 \\
\midrule
Replicated2k  & 3 & $\mathbb{Z}_{2^k}$ & HM/SH  & Server (replicated)  & 313.1  & 53.9     & 835.6   & 53.8     & 1148.8 \\
Semi2k        & 3 & $\mathbb{Z}_{2^k}$ & DM/SH  & Server (additive)    & 528.0  & 31060.2  & 841.4   & 161.5    & 1369.5 \\
PSReplicated2k & 3 & $\mathbb{Z}_{2^k}$ & HM/Mal & Server (replicated)  & 324.2  & 266.3    & 1197.1  & 132.9    & 1521.5 \\
SPDZ2k        & 3 & $\mathbb{Z}_{2^k}$ & DM/Mal & Server (additive)    & 3165.9 & 285821.0 & 3067.3  & 242.2    & 6233.8 \\
\bottomrule
\end{tabular}
\end{table*}

Table \ref{tab:protocol-results} shows the overhead of our MPC solution for performing secure federated computations. We follow the setup of the previous experiments on GTSRB. We combine the analysis for both FL averaging and guidance matrix computation as their core secure operation is identical: a federated average across client parameters.
The final normalization step required for the guidance matrix is handled locally by each client to minimize MPC overhead.
  We report offline and online phases separately and compare six protocol configurations.

A significant factor in MPC overhead for a large model such as 2CNN (${\sim}1.7$M parameters) is input supply, i.e., generation and distribution of secret-shared inputs from clients to parties. In MP-SPDZ's outsourced client protocol~\cite{ExternalIO},  servers generate a shared random mask~$[r]$, reveal it to the client,
who returns the masked input $x + r$, from which servers compute $[x] = (x{+}r) - [r]$. For malicious protocols, each input value additionally consumes a multiplication triple to authenticate the mask~\cite{ddnnt15}. To address this, we extended the MP-SPDZ \texttt{ExternalIO} interface with \textit{client-side} share generation for semi-honest protocols: clients directly compute and distribute secret shares
eliminating server-side preprocessing. As shown in Table~\ref{tab:protocol-results}, this reduces total time from $1148.8$s to $20.5$s for Replicated2k. We note that even when server-side masking is required (e.g., for malicious protocols), the preprocessing is offline and \textit{input-independent}. For example, an OEM can pre-distribute the randomness to vehicles in advance.

The choice of
MPC protocol also significantly impacts overhead. With server-side input masking, total execution time ranges from $1148.8$s (Replicated2k) to $6233.8$s (SPDZ2k). SPDZ2k incurs the highest cost due to malicious security under a dishonest majority, requiring MAC-authenticated preprocessing that generates over 285~GB in the offline phase. PSReplicated2k~\cite{prss} provides malicious security under an honest majority at moderate cost. Semi2k and Replicated2k assume semi-honest behavior and are most efficient, finishing within ${\sim}30$s with client-side share generation.
Finally, Table~\ref{tab:scaling-results} shows  scaling  with the number of clients: Replicated2k setting, which requires no preprocessing, scales more efficiently, while  PSReplicated2k setting with server-side masking shows  noticeable overhead per each additional client.

For fleet-learning deployments, carefully designed MPC
protocols can keep overhead practical: semi-honest execution with client-side share generation completes under 30s with negligible preprocessing, while the higher
offline cost under malicious security can be scheduled
during vehicle downtime.

\begin{table}[t]
\centering
\caption{Scaling of MPC overhead with the number of FL clients.}
\label{tab:scaling-results}
\renewcommand{\arraystretch}{1.12}
\begin{tabular}{l r | r r | r}
\toprule
\textbf{Protocol} & \textbf{\# Clients}
& \textbf{Offline (s)} & \textbf{Online (s)} & \textbf{Total (s)} \\
\midrule
Replicated2k   &  2 & $\approx$0 &  20.4 &  20.5 \\
Replicated2k   &  4 & $\approx$0 &  37.9 &  37.9 \\
Replicated2k   & 10 & $\approx$0 &  75.6 &  75.7 \\
\midrule
PSReplicated2k &  2 &  324.2 & 1197.1 & 1521.5 \\
PSReplicated2k &  4 & 1006.2 & 1544.8 & 2551.2 \\
PSReplicated2k & 10 & 3179.4 & 2624.1 & 5803.8 \\
\bottomrule
\end{tabular}
\end{table}


\section{Conclusion}
\label{conc}

We investigated the privacy risks that arise when 
non-IID FL  optimization approaches share additional gradient information, focusing on  \texttt{FedLEx} in the traffic sign classification domain. Our analysis shows that sharing gradient deviations enables a malicious server to infer client-specific label distributions
 with 50\%--84\% accuracy
on real traffic-sign datasets.
To address this concern, we proposed \texttt{PPFLex} that replaces single-server aggregation with a three-server MPC architecture that securely computes global guidance and model aggregation. We evaluated multiple ring-based protocols and quantified runtime and communication trade-offs under different security assumptions.
These findings emphasize the need for optimized protocols and tailored MPC implementations to improve efficiency, extending scalability to larger models and more clients.

\bibliographystyle{IEEEtran}
\bibliography{bibliography}

\begin{thebibliography}{10}
\providecommand{\url}[1]{#1}
\csname url@samestyle\endcsname
\providecommand{\newblock}{\relax}
\providecommand{\bibinfo}[2]{#2}
\providecommand{\BIBentrySTDinterwordspacing}{\spaceskip=0pt\relax}
\providecommand{\BIBentryALTinterwordstretchfactor}{4}
\providecommand{\BIBentryALTinterwordspacing}{\spaceskip=\fontdimen2\font plus
\BIBentryALTinterwordstretchfactor\fontdimen3\font minus \fontdimen4\font\relax}
\providecommand{\BIBforeignlanguage}[2]{{%
\expandafter\ifx\csname l@#1\endcsname\relax
\typeout{** WARNING: IEEEtran.bst: No hyphenation pattern has been}%
\typeout{** loaded for the language `#1'. Using the pattern for}%
\typeout{** the default language instead.}%
\else
\language=\csname l@#1\endcsname
\fi
#2}}
\providecommand{\BIBdecl}{\relax}
\BIBdecl

\bibitem{Kou2024pFedLVMAL}
W.-B. Kou, Q.~Lin, M.~Tang, S.~Xu, R.~Ye, Y.~Leng, S.~Wang, G.~Li, Z.~Chen, G.~Zhu \emph{et~al.}, ``pfedlvm: A large vision model (lvm)-driven and latent feature-based personalized federated learning framework in autonomous driving,'' \emph{IEEE Transactions on Intelligent Transportation Systems}, 2025.

\bibitem{Chellapandi2023FederatedLF}
V.~P. Chellapandi, L.~Yuan, C.~G. Brinton, S.~H. {\.Z}ak, and Z.~Wang, ``Federated learning for connected and automated vehicles: A survey of existing approaches and challenges,'' \emph{IEEE Transactions on Intelligent Vehicles}, vol.~9, no.~1, pp. 119--137, 2023.

\bibitem{yuca2025survey}
N.~Yuca, N.~Matyunin, E.~Arzoglou, N.~A. Anagnostopoulos, and S.~Katzenbeisser, ``A survey on privacy-preserving computing in the automotive domain,'' \emph{ACM Computing Surveys}, vol.~58, no.~5, pp. 1--36, 2025.

\bibitem{Nidamanuri2022APR}
J.~Nidamanuri, C.~Nibhanupudi, R.~Assfalg, and H.~Venkataraman, ``A progressive review: Emerging technologies for adas driven solutions,'' \emph{IEEE Transactions on Intelligent Vehicles}, vol.~7, no.~2, pp. 326--341, 2021.

\bibitem{kesbhat2025TrafficSD}
H.~Kesbhat, N.~Jetty, M.~Murmu, L.~Vignesh, and K.~Himabindu, ``Traffic sign detection, classification, and violation logging system for intelligent driver assistance,'' \emph{International Journal on Science and Technology}, 2025.

\bibitem{Ertler2019TheMT}
C.~Ertler, J.~Mislej, T.~Ollmann, L.~Porzi, G.~Neuhold, and Y.~Kuang, ``The mapillary traffic sign dataset for detection and classification on a global scale,'' in \emph{European conference on computer vision}.\hskip 1em plus 0.5em minus 0.4em\relax Springer, 2020, pp. 68--84.

\bibitem{Hossein2025ArgosAD}
S.~M. H.~S. Hossein, A.~Hosseini, S.~H. Manesh, and A.~Shahriary, ``Argos: A decentralized federated system for detection of traffic signs in cavs,'' \emph{arXiv preprint arXiv:2508.12712}, 2025.

\bibitem{Liang2024FederatedLW}
J.~Liang, J.~Li, J.~Zhang, and T.~Zang, ``Federated learning with data-free distillation for heterogeneity-aware autonomous driving,'' in \emph{2024 International Joint Conference on Neural Networks (IJCNN)}.\hskip 1em plus 0.5em minus 0.4em\relax IEEE, 2024, pp. 1--7.

\bibitem{Padaria2023TrafficSC}
A.~A. Padaria, A.~A. Mehta, N.~K. Jadav, S.~Tanwar, D.~Garg, A.~Singh, G.~Pau, and G.~Sharma, ``Traffic sign classification for autonomous vehicles using split and federated learning underlying 5g,'' \emph{IEEE Open Journal of Vehicular Technology}, vol.~4, pp. 877--892, 2023.

\bibitem{Truong2020PrivacyPI}
N.~Truong, K.~Sun, S.~Wang, F.~Guitton, and Y.~Guo, ``Privacy preservation in federated learning: An insightful survey from the gdpr perspective,'' \emph{Computers \& Security}, vol. 110, p. 102402, 2021.

\bibitem{zhu2021federated}
H.~Zhu, J.~Xu, S.~Liu, and Y.~Jin, ``Federated learning on non-iid data: A survey,'' \emph{Neurocomputing}, vol. 465, pp. 371--390, 2021.

\bibitem{Lian2024TrafficSR}
Z.~Lian, Q.~Zeng, W.~Wang, D.~Xu, W.~Meng, and C.~Su, ``Traffic sign recognition using optimized federated learning in internet of vehicles,'' \emph{IEEE internet of things journal}, vol.~11, no.~4, pp. 6722--6729, 2023.

\bibitem{McMahan2016CommunicationEfficientLO}
B.~McMahan, E.~Moore, D.~Ramage, S.~Hampson, and B.~A. y~Arcas, ``Communication-efficient learning of deep networks from decentralized data,'' in \emph{Artificial intelligence and statistics}.\hskip 1em plus 0.5em minus 0.4em\relax Pmlr, 2017, pp. 1273--1282.

\bibitem{Collins2022FedAvgWF}
L.~Collins, H.~Hassani, A.~Mokhtari, and S.~Shakkottai, ``Fedavg with fine tuning: Local updates lead to representation learning,'' \emph{Advances in Neural Information Processing Systems}, vol.~35, pp. 10\,572--10\,586, 2022.

\bibitem{Shao2023ASO}
J.~Shao, Z.~Li, W.~Sun, T.~Zhou, Y.~Sun, L.~Liu, Z.~Lin, Y.~Mao, and J.~Zhang, ``A survey of what to share in federated learning: Perspectives on model utility, privacy leakage, and communication efficiency,'' \emph{arXiv preprint arXiv:2307.10655}, 2023.

\bibitem{FedLEx}
C.~Intern{\`o}, M.~Olhofer, Y.~Jin, and B.~Hammer, ``Federated loss exploration for improved convergence on non-iid data,'' in \emph{2024 International Joint Conference on Neural Networks (IJCNN)}.\hskip 1em plus 0.5em minus 0.4em\relax IEEE, 2024, pp. 1--8.

\bibitem{Chai2023ASF}
D.~Chai, L.~Wang, L.~Yang, J.~Zhang, K.~Chen, and Q.~Yang, ``A survey for federated learning evaluations: Goals and measures,'' \emph{IEEE transactions on knowledge and data engineering}, vol.~36, no.~10, pp. 5007--5024, 2024.

\bibitem{Lyu2020PrivacyAR}
L.~Lyu, H.~Yu, X.~Ma, C.~Chen, L.~Sun, J.~Zhao, Q.~Yang, and P.~S. Yu, ``Privacy and robustness in federated learning: Attacks and defenses,'' \emph{IEEE transactions on neural networks and learning systems}, vol.~35, no.~7, pp. 8726--8746, 2022.

\bibitem{Mothukuri2021ASO}
V.~Mothukuri, R.~M. Parizi, S.~Pouriyeh, Y.~Huang, A.~Dehghantanha, and G.~Srivastava, ``A survey on security and privacy of federated learning,'' \emph{Future Generation Computer Systems}, vol. 115, pp. 619--640, 2021.

\bibitem{DeepLF}
L.~Zhu, Z.~Liu, and S.~Han, ``Deep leakage from gradients,'' \emph{Advances in neural information processing systems}, vol.~32, 2019.

\bibitem{Doan2023ASO}
T.~V.~T. Doan, M.-L. Messai, G.~Gavin, and J.~Darmont, ``A survey on implementations of homomorphic encryption schemes: Dt van thao et al.'' \emph{The Journal of Supercomputing}, vol.~79, no.~13, pp. 15\,098--15\,139, 2023.

\bibitem{Ji2014DifferentialPA}
Z.~Ji, Z.~C. Lipton, and C.~Elkan, ``Differential privacy and machine learning: a survey and review,'' \emph{arXiv preprint arXiv:1412.7584}, 2014.

\bibitem{Lindell2020SecureMC}
Y.~Lindell, ``Secure multiparty computation,'' \emph{Communications of the ACM}, vol.~64, no.~1, pp. 86--96, 2020.

\bibitem{GmezSerna2018ClassificationOT}
C.~G. Serna and Y.~Ruichek, ``Classification of traffic signs: The european dataset,'' \emph{IEEe Access}, vol.~6, pp. 78\,136--78\,148, 2018.

\bibitem{gtsrb}
J.~Stallkamp, M.~Schlipsing, J.~Salmen, and C.~Igel, ``The german traffic sign recognition benchmark: a multi-class classification competition,'' in \emph{The 2011 international joint conference on neural networks}.\hskip 1em plus 0.5em minus 0.4em\relax IEEE, 2011, pp. 1453--1460.

\bibitem{belgium}
R.~Timofte, K.~Zimmermann, and L.~Van~Gool, ``Multi-view traffic sign detection, recognition, and 3d localisation,'' \emph{Machine vision and applications}, vol.~25, no.~3, pp. 633--647, 2014.

\bibitem{Fereidooni2021SAFELearnSA}
H.~Fereidooni, S.~Marchal, M.~Miettinen, A.~Mirhoseini, H.~M{\"o}llering, T.~D. Nguyen, P.~Rieger, A.-R. Sadeghi, T.~Schneider, H.~Yalame \emph{et~al.}, ``Safelearn: Secure aggregation for private federated learning,'' in \emph{2021 IEEE security and privacy workshops (SPW)}.\hskip 1em plus 0.5em minus 0.4em\relax IEEE, 2021, pp. 56--62.

\bibitem{Gehlhar2023SafeFLMF}
T.~Gehlhar, F.~Marx, T.~Schneider, A.~Suresh, T.~Wehrle, and H.~Yalame, ``Safefl: Mpc-friendly framework for private and robust federated learning,'' in \emph{2023 IEEE security and privacy workshops (SPW)}.\hskip 1em plus 0.5em minus 0.4em\relax IEEE, 2023, pp. 69--76.

\bibitem{Rathee2023ELSASA}
M.~Rathee, C.~Shen, S.~Wagh, and R.~A. Popa, ``Elsa: Secure aggregation for federated learning with malicious actors,'' in \emph{2023 IEEE Symposium on Security and Privacy (SP)}.\hskip 1em plus 0.5em minus 0.4em\relax IEEE, 2023, pp. 1961--1979.

\bibitem{wwfl}
F.~Marx, T.~Schneider, A.~Suresh, T.~Wehrle, C.~Weinert, and H.~Yalame, ``Ww-fl: Secure and private large-scale federated learning,'' \emph{arXiv preprint arXiv:2302.09904}, 2023.

\bibitem{spefl}
L.~Shen, Z.~Ke, J.~Shi, X.~Zhang, Y.~Sun, J.~Zhao, X.~Wang, and X.~Zhao, ``Spefl: Efficient security and privacy-enhanced federated learning against poisoning attacks,'' \emph{IEEE Internet of Things Journal}, vol.~11, no.~8, pp. 13\,437--13\,451, 2023.

\bibitem{scionfl}
Y.~Ben-Itzhak, H.~M{\"o}llering, B.~Pinkas, T.~Schneider, A.~Suresh, O.~Tkachenko, S.~Vargaftik, C.~Weinert, H.~Yalame, and A.~Yanai, ``Scionfl: Efficient and robust secure quantized aggregation,'' in \emph{2024 IEEE Conference on Secure and Trustworthy Machine Learning (SaTML)}.\hskip 1em plus 0.5em minus 0.4em\relax IEEE, 2024, pp. 490--511.

\bibitem{tan2022federated}
Y.~Tan, G.~Long, J.~Ma, L.~Liu, T.~Zhou, and J.~Jiang, ``Federated learning from pre-trained models: A contrastive learning approach,'' \emph{Advances in neural information processing systems}, vol.~35, pp. 19\,332--19\,344, 2022.

\bibitem{nguyen2022where}
J.~Nguyen, J.~Wang, K.~Malik, M.~Sanjabi, and M.~Rabbat, ``Where to begin? on the impact of pre-training and initialization in federated learning,'' \emph{arXiv preprint arXiv:2206.15387}, 2022.

\bibitem{9207469}
C.~Briggs, Z.~Fan, and P.~Andras, ``Federated learning with hierarchical clustering of local updates to improve training on non-iid data,'' in \emph{2020 international joint conference on neural networks (IJCNN)}.\hskip 1em plus 0.5em minus 0.4em\relax IEEE, 2020, pp. 1--9.

\bibitem{10114536730383673151}
M.~S. Islam, S.~Javaherian, F.~Xu, X.~Yuan, L.~Chen, and N.-F. Tzeng, ``Fedclust: Tackling data heterogeneity in federated learning through weight-driven client clustering,'' in \emph{Proceedings of the 53rd International Conference on Parallel Processing}, 2024, pp. 474--483.

\bibitem{neth2025feddiversetacklingdataheterogeneity}
G.~D. N{\'e}meth, E.~Fan{\`\i}, Y.~J. Ng, B.~Caputo, M.~{\'A}. Lozano, N.~Oliver, and N.~Quadrianto, ``Feddiverse: tackling data heterogeneity in federated learning with diversity-driven client selection,'' in \emph{2025 3rd International Conference on Federated Learning Technologies and Applications (FLTA)}.\hskip 1em plus 0.5em minus 0.4em\relax IEEE, 2025, pp. 432--440.

\bibitem{loshchilov2019decoupledweightdecayregularization}
I.~Loshchilov and F.~Hutter, ``Decoupled weight decay regularization,'' \emph{arXiv preprint arXiv:1711.05101}, 2017.

\bibitem{wu2018group}
Y.~Wu and K.~He, ``Group normalization,'' in \emph{Proceedings of the European conference on computer vision (ECCV)}, 2018, pp. 3--19.

\bibitem{ruiz2025group}
J.~Zhong, H.-Y. Chen, and W.-L. Chao, ``Making batch normalization great in federated deep learning,'' \emph{arXiv preprint arXiv:2303.06530}, 2023.

\bibitem{Stergiou2022AFL}
K.~D. Stergiou, K.~E. Psannis, V.~Vitsas, and Y.~Ishibashi, ``A federated learning approach for enhancing autonomous vehicles image recognition,'' in \emph{2022 4th International Conference on Computer Communication and the Internet (ICCCI)}.\hskip 1em plus 0.5em minus 0.4em\relax IEEE, 2022, pp. 87--90.

\bibitem{Li2019FederatedLC}
T.~Li, A.~K. Sahu, A.~Talwalkar, and V.~Smith, ``Federated learning: Challenges, methods, and future directions,'' \emph{IEEE signal processing magazine}, vol.~37, no.~3, pp. 50--60, 2020.

\bibitem{Kingma2014AdamAM}
D.~P. Kingma and J.~Ba, ``Adam: A method for stochastic optimization,'' \emph{arXiv preprint arXiv:1412.6980}, 2014.

\bibitem{nikolic_guided_2023}
D.~Nikoli{\'c}, D.~Andri{\'c}, and V.~Nikoli{\'c}, ``Guided transfer learning,'' \emph{arXiv preprint arXiv:2303.16154}, 2023.

\bibitem{mp-spdz}
M.~Keller, ``{MP-SPDZ}: A versatile framework for multi-party computation,'' in \emph{Proceedings of the 2020 ACM SIGSAC Conference on Computer and Communications Security}, 2020.

\bibitem{ExternalIO}
\BIBentryALTinterwordspacing
{MP-SPDZ}, ``{ExternalIO - MP-SPDZ},'' 2025, accessed: 14-Feb-2025. [Online]. Available: \url{https://github.com/data61/MP-SPDZ/tree/master/ExternalIO}
\BIBentrySTDinterwordspacing

\bibitem{ddnnt15}
I.~Damg{\aa}rd, K.~Damg{\aa}rd, K.~Nielsen, P.~S. Nordholt, and T.~Toft, ``Confidential benchmarking based on multiparty computation,'' in \emph{International Conference on Financial Cryptography and Data Security}.\hskip 1em plus 0.5em minus 0.4em\relax Springer, 2016, pp. 169--187.

\bibitem{prss}
H.~Eerikson, M.~Keller, C.~Orlandi, P.~Pullonen, J.~Puura, and M.~Simkin, ``Use your brain! arithmetic 3pc for any modulus with active security,'' \emph{Cryptology ePrint Archive}, 2019.

\end{thebibliography}

\end{document}